\documentclass{article}
\usepackage{spconf,amsmath,graphicx,bbm}
\usepackage{multirow, hhline, stmaryrd}

\title{Adversarial Teacher-Student Learning for Unsupervised Domain Adaptation}
\name{Zhong Meng$^{1,2}$\sthanks{Zhong Meng performed the work while he was a research intern
	at Microsoft AI and Research, Redmond, WA, USA.}, Jinyu Li$^{1}$, Yifan Gong$^{1}$, Biing-Hwang (Fred) Juang$^{2}$
}
\address{$^{1}$ Microsoft AI and Research, Redmond, WA, USA
\\ $^{2}$ Georgia Institute of Technology, Atlanta, GA, USA
}

\begin{document}
\ninept
\maketitle

\begin{abstract}


The teacher-student (T/S) learning has been shown effective in unsupervised domain adaptation \cite{ts_adapt}. It is a form of transfer learning, not in terms of the transfer of recognition decisions, but the knowledge of posteriori probabilities in the source domain as evaluated by the teacher model. It learns to handle the speaker and environment variability inherent in and restricted to the speech signal in the target domain without proactively addressing the robustness to other likely conditions. Performance degradation may thus ensue. In this work, we advance T/S learning by proposing adversarial T/S learning to explicitly achieve condition-robust unsupervised domain adaptation. In this method, a student acoustic model and a condition classifier are jointly optimized to minimize the Kullback-Leibler divergence between the output distributions of the teacher and student models, and simultaneously, to min-maximize the condition classification loss. A condition-invariant deep feature is learned in the adapted student model through this procedure. We further propose multi-factorial adversarial T/S learning which suppresses condition variabilities caused by multiple factors simultaneously. Evaluated with the noisy CHiME-3 test set, the proposed methods achieve relative word error rate improvements of 44.60\% and 5.38\%, respectively, over a clean source model and a strong T/S learning baseline model.





%

\end{abstract}
\begin{keywords}
teacher-student learning, adversarial training, domain adaptation, parallel
unlabeled data
\end{keywords}

\section{Introduction}
\label{sec:intro}
With the advance of deep learning, the performance of automatic speech
recognition (ASR) has been greatly improved \cite{ 
sainath2011making, jaitly2012application, DNN4ASR-hinton2012,
deng2013recent, yu2017recent}. However, the ASR still suffers from
large performance degradation when a well-trained acoustic model is 
presented in a new domain \cite{Li14overview, Li15robust}. 
Many domain adaptation techniques were proposed to address this issue, such
as regularization-based \cite{kld_yu, map_huang, l2_liao, multi_huang},
transformation-based \cite{feature_seide, lhuc_pawel_1, zhao2015investigating}, singular value
decomposition-based \cite{svd_xue_1,svd_xue_2, svd_zhao} and subspace-based
\cite{ivector_saon, sc_xue, fhl, zhao2017extended} approaches.  Although these methods
effectively mitigate the mismatch between source and target domains, they
reply on the transcription or the first-pass decoding hypotheses of the adaptation
data.

To address these limitations, teacher-student (T/S) learning
\cite{ts_learning} is used to achieve unsupervised
adaptation \cite{ts_adapt} with no exposure to any transcription or decoded
hypotheses of the adaptation data. In T/S learning, the  posteriors
generated by the teacher model are used in lieu of the hard labels derived
from the transcriptions to train the target-domain student model.  Although
T/S learning achieves large word error rate (WER) reduction in domain adaptation, 
it is similar to the traditional
training criterion such as cross entropy (CE) which implicitly handles the variations in each speech
unit (e.g.  senone) caused by the speaker and environment variability in
addition to phonetic variations. 

Recently, adversarial training has become a hot topic in deep learning with
its great success in estimating generative models \cite{gan}. It has also
been applied to noise-robust \cite{grl_shinohara, grl_serdyuk, grl_sun, dsn_meng} and speaker-invariant 
\cite{meng2018speaker} ASR using gradient reversal layer \cite{grl_ganin} or domain separation network \cite{dsn}.  A deep
intermediate feature is learned to be both discriminative for the main task
of senone classification and invariant with respect to the shifts among
different conditions. Here, one condition refers to one particular speaker
or one acoustic environment. 
For unsupervised adaptation, both the T/S learning and adversarial training forgo the need for
any labels or decoded results of the adaptation data.  T/S learning is more
suitable for the situation where parallel data is available since the paired data allows the 
student model to be better-guided by the knowledge from the source model, while adversarial
training is more powerful when such data is not available. 

To benefit from both methods, in this work, we advance T/S learning with \emph{adversarial
T/S training} for condition-robust unsupervised domain adaptation, where a
student acoustic model and a domain classifier are jointly trained to
minimize the Kullback-Leibler (KL) divergence between the output
distributions of the teacher and student
models as well as to min-maximize the condition classification loss through
adversarial multi-task learning.  A senone-discriminative and \emph{condition-invariant} 
deep feature is learned in the adapted student model through this procedure. 
Based on this, we further propose the 
\emph{multi-factorial adversarial (MFA)} T/S learning where the condition 
variabilities caused by multiple factors are minimized simultaneously.
Evaluated with the noisy CHiME-3 test set, the proposed method
achieves 44.60\% and 5.38\% relative WER improvements
over the clean model and a strong T/S adapted baseline acoustic model, respectively.




\section{Teacher-Student Learning}
\label{sec:ts}

By using T/S learning for unsupervised adaption, we want to learn a student
acoustic model that can accurately predict the senone posteriors of the
target-domain data from a well-trained source-domain teacher acoustic
model. To achieve this, we only need two sequences of \emph{unlabeled} parallel
data, i.e., an input sequence of  source-domain speech frames to the teacher model
$X^T=\{x^T_{1}, \ldots, x^T_{N}\}$ and an input sequence of target-domain speech frames to the
 student model $X^S=\{x^S_{1}, \ldots, x^S_{N}\}$.  $X^T$ and
$X^S$ are parallel to each other, i.e, each pair of $x^S_i$ and $x^T_i,
\forall i \in \{1, \ldots, N\}$ are frame-by-frame synchronized.

T/S learning aims at minimizing the Kullback-Leibler (KL) divergence
between the output distributions of the teacher model and the student model
by taking the unlabeled parrallel data $X^T$ and $X^S$ as the input to the
models.  The KL divergence between the teacher and student output
distributions $p_T(q|x^T_i; \theta_T)$ and $p_S(q|x^S_i; \theta_S)$ is 
\begin{align}
	\mathcal{KL}(p_T||p_S) = \sum_{i}\sum_{q\in \mathcal{Q}} p_T(q|x^T_i; \theta_T) \log
	\left(\frac{p_T(q|x^T_i; \theta_T)}{p_S(q|x^S_i; \theta_S)} 
	\right) 
	\label{eqn:kld_ts}
\end{align}
where $q$ is one of the senones in the senone set $\mathcal{Q}$, $i$ is the frame
index, $\theta_T$ and $\theta_S$ are the parameters of the teacher and
student models respectively.
To learn a student network that approximates the given teacher network, we
minimize the KL divergence with respect to only the parameters of the
student network while keeping the parameters of the teacher model fixed,
which is equivalent to minimizing the loss function below:
\begin{align}
	\mathcal{L}(\theta_S) = -\sum_i\sum_{q\in \mathcal{Q}}
	p_T(q|x^T_i;\theta_T) \log p_S(q|x^S_i;\theta_S)
	\label{eqn:loss_ts}
\end{align}

The target domain data used to adapt the student model
is usually recorded under multiple conditions, i.e., the adaptation data
often comes from a large number of different talkers speaking under various
types of environments (e.g., home, bus, restaurant and etc).  T/S
learning can only implicitly handle the inherent speaker and environment variability in
the speech signal and its robustness can be improved if it can explicitly handle
the condition invariance.



\section{Adversarial Teacher-Student Learning}
\label{sec:ats}
\begin{figure}[htpb!]
	\centering
	\includegraphics[width=1.0\columnwidth]{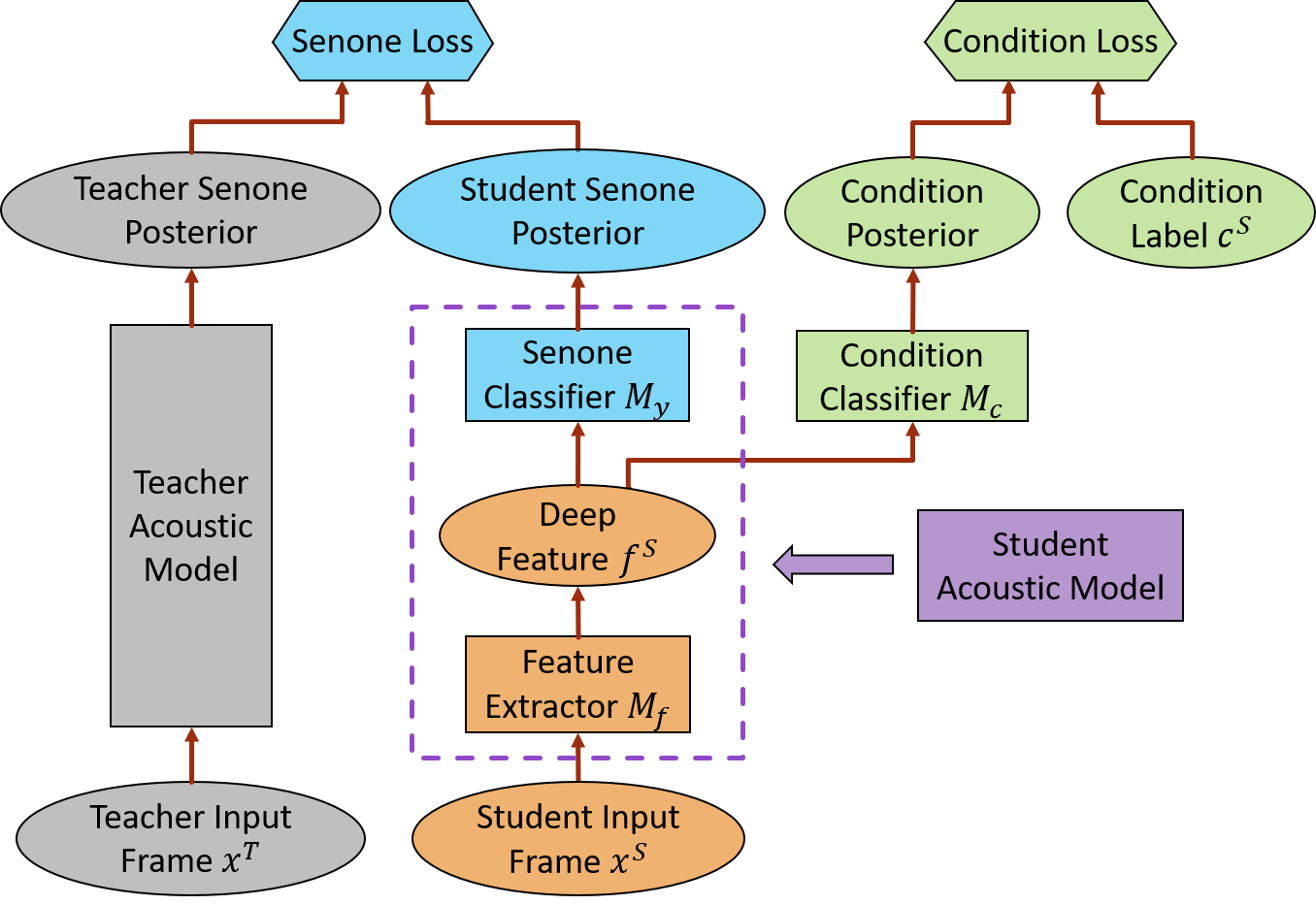}
	\caption{The framework of adversarial T/S learning for
	unsupervised adaptation of the acoustic models}
	\label{fig:ats}
\end{figure}

In this section, we propose the \emph{adversarial T/S learning} (see Fig.
\ref{fig:ats}) to effectively suppress the condition (i.e., speaker and
environment) variations in the speech signal and achieve robust
unsupervised adaptation with multi-conditional adaptation data.

Similar to the T/S learning, we first clone the student acoustic model from
the teacher and use unlabeled parallel data as the input to adapt the
student model. To achieve condition-robustness, we learn a
\emph{condition-invariant} and \emph{senone-discriminative} deep feature in
the adapted student model through the senone posteriors generated by the
teacher model and the condition label for each frame. In order to do so, we
view the first few layers of the 
acoustic model as a
feature extractor 
with parameters $\theta_f$ that maps input
speech frames $X^S$ of different conditions to deep intermediate features
$F^S=\{f_1^S, \ldots, f_N^S\}$ and the upper layers of the student network as a
senone classifier  $M_y$ with parameters $\theta_y$ that maps the intermediate
features $F^S$ to the senone posteriors $p_S(q|f^S_i; \theta_y), q\in
\mathcal{Q}$ as follows:
\begin{align}
	M_y(f_i^S) = M_y(M_f(x^S_i)) = p_S(q | x^S_i; \theta_f,
\theta_y)
	\label{eqn:senone_classify}
\end{align}
where we have $\theta_S = \{\theta_f, \theta_y\}$ as the student model.

We further introduce a condition classifier network $M_c$ with $\theta_c$
which maps the deep features $F^S$ to the condition posteriors $p_c(a |
x^S_i; \theta_c, \theta_f), a \in \mathcal{A}$ as follows:
\begin{align}
	M_c(M_f(x^S_i)) & = p_c(a | x^S_i; \theta_c, \theta_f)
	\label{eqn:domain_classify}
\end{align}
where $a$ is one condition in the set of all conditions $\mathcal{A}$. 

To make the deep features $F^S$ condition-invariant, the distributions of
the features from different conditions should be as close to each other as
possile. Therefore, the $M_f$ and $M_c$ are jointly trained with an
adversarial objective, in which $\theta_f$ is adjusted to \emph{maximize}
the condition classification loss
$\mathcal{L}_{\text{condition}}(\theta_f, \theta_c)$ while $\theta_c$ is adjusted
to \emph{minimize} the $\mathcal{L}_{\text{condition}}(\theta_f, \theta_c)$ below:
\begin{align}
	& \mathcal{L}_{\text{condition}}(\theta_f, \theta_c) = - \sum_{i}^{N} \log
	p_c(c^S_i | x^S_i; \theta_f, \theta_c)\nonumber \\
	& \quad \quad \quad \quad \quad \quad = - \sum_{i}^{N} \sum_{a\in \mathcal{A}} \mathbbm{1}_{[a =
	c^S_i]} \log M_c(M_f(x^S_i)) \label{eqn:loss_cond1}
\end{align}
where $c^S_i$ denote the condition label for the input frame
$x^S_i$ of the student model. 


This minimax competition will first increase the discriminativity of 
$M_c$ and the condition-invariance of the features generated by $M_f$ and
will eventually converge to the point where $M_f$ generates extremely
confusing features that $M_c$ is unable to distinguish.

At the same time, we use T/S learning to let the behavior of the student model
in the target domain approach the behavior of  the  teacher
model in the source domain by minimizing the KL divergence
of the output distributions between the student and teacher acoustic
models. Equivalently, we minimize the loss function in Eq.
\eqref{eqn:loss_ts} as re-formulated below:
\begin{align}
	& \mathcal{L}_{\text{TS}}(\theta_f, \theta_y) = -\sum_i\sum_{q\in \mathcal{Q}}
	p_T(q|x^T_i;\theta_f, \theta_y) M_y(M_f(x^S_i))
	\label{eqn:loss_senone}
\end{align}
In adversarial T/S learning, the student network and the condition classifier
network are trained to jointly optimize the primary task of T/S learning
using soft targets from the teacher model and the secondary task of
condition classification with an adversarial objective function. Therefore,
the total loss is constructed as
\begin{align}
	&\mathcal{L}_{\text{total}}(\theta_f, \theta_y, \theta_c) =
	\mathcal{L}_{\text{TS}}(\theta_f, \theta_y)
	- \lambda \mathcal{L}_{\text{condition}}(\theta_f, \theta_c)
	\label{eqn:loss_total}
\end{align}
where $\lambda$ controls the trade-off between the T/S loss and the condition classification loss
 in Eq.\eqref{eqn:loss_senone} and Eq.\eqref{eqn:loss_cond1} respectively.
 
We need to find the optimal parameters $\hat{\theta}_y, \hat{\theta}_f$ and $\hat{\theta}_c$ such that
\begin{align}
    (\hat{\theta}_f, \hat{\theta}_y) = \min_{\theta_y, \theta_f} \mathcal{L}_{\text{total}}(\theta_f, \theta_y, \hat{\theta}_c) \label{eqn:min_fy} \\
    \hat{\theta}_c = \max_{\theta_c} \mathcal{L}_{\text{total}}(\hat{\theta}_f, \hat{\theta}_y, \theta_c)
    \label{eqn:max_c}
\end{align}

The parameters are updated as follows via back propagation through time
with stochastic gradient descent (SGD):
\begin{align}
	& \theta_f \leftarrow \theta_f - \mu \left[ \frac{\partial
		\mathcal{L}_{\text{TS}}}{\partial \theta_f} - \lambda \frac{\partial
			\mathcal{L}_{\text{condition}}}{\partial
			\theta_f}
		\right]
		\label{eqn:grad_f} \\
	& \theta_c \leftarrow \theta_c - \mu \frac{\partial
		\mathcal{L}_{\text{condition}}}{\partial \theta_c} \label{eqn:grad_c} \\
	& \theta_y \leftarrow \theta_y - \mu \frac{\partial
		\mathcal{L}_{\text{TS}}}{\partial \theta_y}
	\label{eqn:grad_y}
\end{align}
where $\mu$ is the learning rate.

Note that the negative coefficient $-\lambda$ in Eq. \eqref{eqn:grad_f}
induces reversed gradient that maximizes
$\mathcal{L}_{\text{condition}}(\theta_f, \theta_c)$ in Eq.  \eqref{eqn:loss_cond1}
and makes the deep feature condition-invariant. 
For easy implementation, gradient reversal layer is introduced in
\cite{grl_ganin}, which acts as an identity transform in the forward propagation
and multiplies the gradient by $-\lambda$ during the backward propagation.

The optimized student network consisting of $M_f$ and $M_y$ is used as the
adapted acoustic model for ASR in the target-domain.

\section{Multi-factorial Adversarial Teacher-Student Learning}
Speaker and environment are two different
factors that contribute to the inherent variability of the speech signal. In
Section \ref{sec:ats}, adversarial T/S learning is proposed to reduce
the variations induced by the single condition. 
For a more comprehensive
and thorough solution to the condition variability problem, we further propose the
multi-factorial adversarial (MFA) T/S learning, in which multiple
factors causing the condition variability are suppressed simultaneously
through adversarial multi-task learning.

In MFA T/S framework, we keep the senone
classifier $M_y$ and feature extractor $M_f$ the same as in adversarial T/S, but
introduce $R$ condition classifiers $M_c^r, r = 1, \ldots, R$. $M_c^r$ maps the 
deep feature to the condition posteriors of factor $r$. To make the deep 
features $F^S$ condition-invariant to each factor, we jointly train $M_f$ and $M_c^r, r = 1, \ldots, R$ with an adversarial objective, in which $\theta_f$ is adjusted to \emph{maximize}
the total condition classification loss of all factors while $\theta_c^r$ is adjusted
to \emph{minimize} the total condition classification loss of all factors. At the 
same time, we minimize the KL divergence between the output distributions of the
teacher and student models. The total 
loss function for MFA T/S learning is formulated as
\begin{align}
  	&\mathcal{L}_{\text{total}}(\theta_f, \theta_y, \theta_c^1, \ldots, \theta_c^R) =
	\mathcal{L}_{\text{TS}}(\theta_f, \theta_y) - 
	\lambda\sum_{r=1}^R \mathcal{L}^r_{\text{condition}}(\theta_c^r, \theta_f) \nonumber \\
	\label{eqn:loss_total}
\end{align}
where $\mathcal{L}_{\text{TS}}$ is defined in Eq. \eqref{eqn:loss_senone} and $\mathcal{L}^{r}_{\text{condition}}$ 
for each $r$ are formulated in the same way as in Eq. \eqref{eqn:loss_cond1}. All the parameters are optimized in the same way as in
Eq. \eqref{eqn:min_fy} to Eq. \eqref{eqn:grad_y}. 
Note that  better performance may be obtained when the condition losses 
have different combination weights. However, we just equally add them together 
in Eq. \eqref{eqn:loss_total} to avoid tuning.

\section{Experiments}
\label{sec:format}

To compare directly with the results in \cite{ts_adapt}, we use exactly the 
same experiment setup as in \cite{ts_adapt}.
We perform unsupervised  adaptation of a clean
long short-term memory (LSTM)- recurrent neural networks (RNN) \cite{sak2014long, meng2017deep, erdogan2016multi} 
acoustic model trained with 375  hours of Microsoft Cortana 
voice assistant data to the noisy CHiME-3
dataset \cite{chime3_barker} using T/S and adversarial T/S learning. The CHiME-3 dataset
incorporates Wall Street Journal (WSJ) corpus sentences spoken in
challenging noisy environments, recorded using a 6-channel tablet. 
The real far-field noisy speech from the 5th microphone channel in
CHiME-3 development data set is used for testing. A standard WSJ 5K word 3-gram
language model (LM) is used for decoding.

The clean acoustic model is an LSTM-RNN trained with cross-entropy
criterion.  We extract 80-dimensional
input log Mel filterbank feature as the input to the acoustic model.  The
LSTM has 4 hidden layers with 1024 units in each layer. A 512-dimensional
projection layer is inserted on top each hidden layer to reduce the number
of parameters. The output layer has 5976 output units predicting senone
posteriors. A WER of 23.16\% is achieved when evaluating the clean model 
on the test data. The clean acoustic model is used as the teacher model in the
following experiments.

\subsection{T/S Learning for Unsupervised Adaptation}
\label{sec:exp_ts}

We first use parallel data consisting of 9137 pairs of
clean and noisy utterances in the CHiME-3 training set (named as
``clean-noisy'') as the adaptation data for T/S learning. In order to let the 
student model be invariant to environments, the training data for student model
should include both clean and noisy data. Therefore, We extend the original T/S
learning work in \cite{ts_adapt} by also including 9137 pairs of the clean and clean utterances in CHiME-3 (named as
``clean-clean'') for adaptation. By perform T/S learning with both the ``clean-noisy'' and ``clean-clean'' parallel data, the learned student model should perform well on both the clean and noisy data because it will approach the behavior of teacher model on clean data no matter it is presented with clean or noisy data. 

The unadapted Cortana model has 6.96\% WER on the clean test set. After T/S learning with both the ``clean-noisy'' and ``clean-clean'' parallel data, the student model has 6.99\% WER on the clean test. As the focus of this study is to improve T/S adaptation on noisy test data, we will only report results with the CHiME-3 real noisy channel 5 test set. 
The WER results on the noisy channel 5  test set of T/S learning are shown in
Table \ref{table:wer_ts}. The T/S learning achieves 13.88\% and 13.56\% average WERs
when adapted to ``clean-noisy'' and ``clean-noisy \& clean-clean''
respectively, which are 40.05\% and 41.45\% relative improvements over
the unadapted clean model.  Note that our experimental setup does
not achieve the state-of-the-art performance on CHiME-3 dataset (e.g., we 
did not perform beamforming, sequence training or use RNN LM for decoding.) since our goal is to simply verify 
the effectiveness of adversarial T/S learning in achieving condition-robust 
unsupervised adaptation.

\begin{table*}[t]
\centering
\begin{tabular}[c]{c|c|c|c|c|c|c}
	\hline
	\hline
	System & Adaptation Data & BUS & CAF & PED & STR & Avg. \\
	\hline
	Unadapted & - & 27.93 & 24.93 & 18.53 & 21.38 & 23.16 \\
	\hline
	\multirow{2}{*}{\begin{tabular}{@{}c@{}} T/S 
		\end{tabular}} & clean-noisy & 16.00 & 15.24 & 11.27 & 13.07 & 13.88 \\
	\hhline{~------}
	& clean-noisy, clean-clean & 15.96 & 14.32 & 11.00 & 13.04 & 13.56 \\
	\hline
	\hline
	\end{tabular}
  \caption{The WER (\%) performance of unadapted, T/S learning adapted LSTM acoustic models for robust ASR on the real noisy channel 5 test set of CHiME-3.}
\label{table:wer_ts}
\end{table*}

\begin{table*}[t]
\centering
\begin{tabular}[c]{c|c|c|c|c|c|c}
	\hline
	\hline
	System & Conditions & BUS & CAF & PED & STR & Avg. \\
	\hline
	 \multirow{4}{*}{\begin{tabular}{@{}c@{}} Adversarial \\ T/S 
		\end{tabular}}& 2 environments & 15.24 & 13.95 & 10.71 &
		12.76 & 13.15  \\
		\hhline{~------}
	& 6 environments& 15.58 & 13.23 & 10.65 &
		13.10 & 13.12  \\
	\hhline{~------}
	& 87 speakers & 14.97 & 13.63 & 10.84 &
		12.24 & 12.90 \\
		\hhline{~------}
	& 87 speakers, 6 environments & 15.38 & 13.08 & 10.47 &
	12.45 & \textbf{12.83} \\
	\hline
	\hline
\end{tabular}
  \caption{The WER (\%) performance of 
	  adversarial T/S learning adapted LSTM acoustic models for robust
	  ASR on the real noisy channel 5 test set of CHiME-3. The adaptation data consists of ``clean-noisy'' and ``clean-clean''.}
\label{table:wer_ats}
\end{table*}




\subsection{Adversarial T/S Learning for Environment-Robust Unsupervised Adaptation}
\label{sec:exp_ats}
We adapt the clean acoustic model with the ``clean-noisy \& clean-clean'' parallel data
using adversarial T/S learning so that the resulting student model is environment invariant. 
The feature extractor $M_f$ is
initialized with the first $N_h$ hidden layers of the clean student LSTM and the
senone classifier $M_y$ is initialized with the last $(4-N_h)$ hidden
layers plus the output layer of the clean LSTM. $N_h$ indicates the position
of the deep feature in the student LSTM. The condition classifier DNN $M_c$ has 2 hidden layers
with 512 units in each hidden layer.

To achieve environment-robust unsupervised adaptation, the condition
classifier DNN $M_c$ is designed to predict the posteriors of different
environments at the output layer. As the adaptation data comes from both
the clean and noisy environments, we first use an $M_c$ with
2 output units to predict these two environments. As shown in Table \ref{table:wer_ats}, the adversarial T/S learning with 2-environment condition classifier achieves
13.15\% WER, which are 43.22\% and 3.02\%
relatively improved over the unadapted and T/S learning adapted models respectively. The $N_h$ and $\lambda$ are fixed at $4$ and $5.0$ respectively in all our experiments.


However, the noisy data in CHiME-3 is recorded under 5 different noisy environments, i.e, on buses
(BUS), in cafes (CAF), in pedestrian areas (PED), at street junctions
(STR) and in booth (BTH). To mitigate the speech variations among these
environments, we further use an $M_c$ with 6 output units to
predict the posteriors of the 5 noisy and 1 clean environments. The WER with 6-environment 
condition classier is 13.12\% which achieves 43.35\% and 3.24\% relative improvement over the
unadapted and T/S learning adapted baseline models respectively. The increasing amount of noisy environments
to be normalized through adversarial T/S learning lead to very limited WER improvement which indicates
that the differences among various kinds of noises are not significant enough in CHiME-3 as compared 
to the distinctions between clean and noisy data.

\vspace{-0.3cm}

\subsection{Adversarial T/S Learning for Speaker-Robust Unsupervised Adaptation}

To achieve speaker-robust unsupervised adaptation, $M_c$ is designed to
predict the posteriors of different speaker identities at the output layer.
The 7138 simulated and 1999 real noisy utterances in CHiME-3 training set
are dictated by 83 and 4 different speakers respectively and the 9137 clean
utterances are read by the same speakers.  In speaker-robust adversarial T/S adaptation, an $M_c$ with 87 output units are used to predict the
posteriors of the 87 speakers. From Table \ref{table:wer_ats}, the adversarial 
T/S learning with 87-speaker condition classifier achieves 12.90\% WER, which is 44.30\% and 4.87\%
relative improvement over the unadapted and T/S adapted baseline models respectively. Larger WER improvement
is achieved by speaker-robust unsupervised adaptation than the environment-robust
methods. This is because T/S learning itself is able to reduce the environment 
variability through directly teaching the noisy student model with the senone 
posteriors from the clean data, which limits the space of improvement that 
environment-robust adversarial T/S learning can obtain.



\subsection{Multi-factorial Adversarial T/S Learning for Unsupervised Adaptation}

Speaker and environment robustness can be achieved simultaneously in unsupervised 
adaptation through MFA T/S learning, in which we need two condition
classifiers: $M_c^1$ 
predicts the posteriors of 87 speakers and $M_c^2$ predicts the posteriors of
1 clean and 5 noisy environments in the adaptation data. From Table \ref{table:wer_ats},
the MFA T/S learning achieves 12.83\% WER, which is 44.60\% and 
5.38\% relative improvement over unadapted and T/S baseline models. The MFA T/S achieves lower WER than all the unifactorial adversarial T/S systems
because it addresses the variations caused by all kinds of factors.



\section{Conclusions}
In this work, adversarial T/S learning is proposed to adapt a clean
acoustic model to highly mismatched multi-conditional noisy data in a purely
unsupervised fashion. To suppress the condition variability in speech signal 
and achieve robust adaptation, a student acoustic model and
a condition classifier are jointly optimized to minimize the KL divergence
between the output distributions of the teacher and student models while
simultaneously mini-maximize condition classification loss.
We further propose the MFA T/S learning where multiple condition classifiers are 
introduced to reduce the condition variabilities caused by different factors.
The proposed methods requires only the unlabeled parallel data for 
domain adaptation.

For environment adaptation on CHiME-3 real noisy channel 5 dataset, T/S learning
gets 41.45\% relative WER reduction from the clean-trained acoustic model. 
Adversarial T/S learning with environment and speaker classifiers achieves 3.24\% and 4.87\% relative WER 
improvements over the strong T/S learning model, respectively. MFA T/S achieves 5.38\% relative 
WER improvement over the same baseline. On top of T/S learning, reducing speaker variability  proves to 
be more effective than reducing environment variability T/S learning on CHiME-3 dataset
because T/S learning already addresses most environment mismatch issues.
Simultaneously decreasing the condition variability in multiple factors can
further slightly improve the ASR performance.

The adversarial T/S learning was verified its effectiveness with a relatively small CHiME-3 task. 
We recently developed a far-field speaker system using thousands of hours data with T/S learning \cite{Li2018Speaker}.  
We are now currently applying the proposed  adversarial T/S learning to further improve our far-field speaker system. 





\vfill\pagebreak
\clearpage

\bibliographystyle{IEEEbib}
\bibliography{strings,refs}

\end{document}